\documentclass[oneside]{article}

\usepackage{graphicx} 
\usepackage{amssymb}
\usepackage{latexsym}
\usepackage{amsmath}
\usepackage{fixltx2e}

\usepackage[sc]{mathpazo} 
\usepackage[T1]{fontenc} 
\usepackage{microtype} 

\usepackage[hmarginratio=1:1,top=32mm,columnsep=20pt]{geometry} 
\usepackage{multicol} 
\usepackage[hang, small,labelfont=bf,up,textfont=it,up]{caption} 
\usepackage{booktabs} 
\usepackage{float} 
\usepackage{hyperref} 

\usepackage{lettrine} 
\usepackage{paralist} 

\usepackage{abstract} 

\usepackage{titlesec} 
\renewcommand\thesection{\Roman{section}} 
\renewcommand\thesubsection{\Roman{subsection}} 
\titleformat{\section}[block]{\large\scshape\centering}{\thesection.}{1em}{} 
\titleformat{\subsection}[block]{\large}{\thesubsection.}{1em}{} 

\usepackage{fancyhdr} 

\title{\vspace{-15mm}\fontsize{24pt}{10pt}\selectfont\textbf{A general potential for molecular dynamics of ion-sputtered surfaces}} 

\author{
\large
\textsc{Akande Raphael O.}\thanks{\normalsize \href{mailto:raphaelakande@myashiritycode.com}{raphaelakande@myashiritycode.com}},\qquad
\textsc{Oyewande Emmanuel O.}\thanks{\normalsize \href{mailto:oe.oyewande@mail1.ui.edu.ng}{oe.oyewande@mail1.ui.edu.ng}}\\
\normalsize Theoretical Physics Group, University of Ibadan \\ 
\vspace{-5mm}
}

\date{}

\begin{document}

\maketitle 

\thispagestyle{fancy} 


\begin{abstract}

\noindent 
\textbf{Erosion of surface atoms of solid materials by ion bombardment (surface-sputtering) causes nano-ripples and quantum dots to self-organise on the surfaces. The self-organisation had been shown, in some sputtering experiments, to be influenced by unexpected contaminants (ions) from vacuum walls. Existing inter-atomic-interaction potentials of Molecular Dynamics (MD) simulations for studying this are unsuitable because they assume two-particle collisions at a time instead of many (including contaminants)-particle collisions (Wider-area Perturbations, (WP)). We designed this study to develop a suitable potential that incorporates WP of the MD. We developed the general potential to account for the possibility of WP due to contaminants (both foreign and local to the material) consequently shifting the eqiulibrium points of the MD the material. For instance, dynamics of Au and Fe were studied with O bombardments/contamination (oxygenated environments), and those of CSiGe were studied with W, Ti, and O. It was found that the phase-shift, of the eqiulibrium points of the MD simulations, for Au stabilised for just five atoms of oxygen while that of Fe did not, when interacting in separate oxygenated environments. A different phase shift was recorded for CSiGe which reflected the usage of different contaminants, under subsequent bombardments with W, Ti and O, in transformations to CSiGeW, CSiGeWTi, and CSiGeWTiO. These results and those for the sampled superconducting materials, such as SrNdCuO and LiFeAs, showed a sensitivity of the general potential, and insensitivity of the existing potentials, to the type and environment of the atoms of these materials. Furthermore, results obtained clearly show that the MD perturbations in a material, via its surface, is peculiar to the surface.}
\end{abstract}



\providecommand{\keywords}[1]{\textbf{keywords~:~}#1}
\keywords{potential, surfaces, sputtering, molecular dynamics perturbation, interactions, energy, dynamics}

\section{Introduction}
\lettrine[nindent=0em,lines=3]{T}he industrial and research understanding of atomic interactions often ignore some pertubations in the molecular dynamics. Some readily available applications, which indicate these perturbations, can be found in the solar panel surface sputtering which is also an important factor in its effectiveness, thus the need for anti-relective coats on its surface \textsuperscript{\cite{1}}. Even the best of water and chemical proof materials \textsuperscript{\cite{2}} and materials that will one day allow us to breathe under water are now being developed \textsuperscript{\cite{3}}. Many atimes there are different degrees of energy absorption on different surfaces. Biological surfaces' energy spread differ from that of the inorganics, this is why it is important to design chemically inert surfaces \textsuperscript{\cite{4}} and develop artificial Antimicrobial Activity of Water Resistant Surface Coating \textsuperscript{\cite{5}}. In some other applications, nano scale energy transport might not be required. However, in others like those above there is a need to study nano scale energy transport so as to effectively monitor the delivery of energy to the active atoms.

\subsection{Problem statement and motivation for research}
 In this work, we want to incoporate the surface sputtering processes \textsuperscript{\cite{6}},\textsuperscript{\cite{7}},\textsuperscript{\cite{8}} to molecular interactions \textsuperscript{\cite{9}}. This is because the materials are not operating in isolation from their environments. While the material is undergoing molecular dynamics (within its bulk), its surface is always being bombarded with foreign particles. In essence, the molecular dynamics is perturbed \textsuperscript{\cite{10}} by the energy thrown downwards by the approching foreign particle. Therefore, all materials are in constant tune with their environment, which implies that we must develop a potential for the molecular interactions to incoporate this surface sputtering section of the interactions. We are convinced that the molecular dynamics perturbations in a material, as a result of bombardments on its surface by foreign particles, is peculiar to the surface. The essence of developing a general potential is to be able to apply the single potential to any (or most) materials located in any environment. In what follows, we discuss some potentials in literature which are not general enough for achieving our goal.

\subsection{Interatomic Potentials for Metallic Systems}
We shall itemize the well known potentials as follows, source: \textsuperscript{\cite{13}}:
\begin{itemize}
\item \begin{equation}H_I^{EAM}=\sum_i F_i[\rho_{h,i}]+\frac{1}{2}\sum_i \sum_{j\ne i}\phi_{ij}\left(r_{ij}\right)\end{equation}
this is the Many-Body Embeded-Atom Model (EAM) Potentials.
where $\rho_{h,i}$ is the electron density of the host at the site of atom $i$,  $F_i[\rho_{h,i}]$ is the embedding functional, that is, the energy to embed an extra atom $i$ into the background electron density,$\rho$ and $\phi_{ij}$ is a pairwise central potential between atoms $i$ and $j$, separated by a distance $r_{ij}$. $\phi_{ij}$ represents the repulsive core-core electrostatic interaction. The electron density of the host lattice is a linear superposition of the individual contributions and is given as:
\begin{equation}\rho_{h,i}=\sum_{j\ne i}\rho_j^*\left(r_{ij}\right)\end{equation}
where $\rho_j^*$ is also a pairwise term and is the electron density of atom $j$ as a function of interatomic separation.
The EAM are developed based on existing Density Functional Theory (DFT). According to DFT, the energy of a collection of atoms can be expressed exactly by a functional of its electronic density.

\item  \begin{equation}H_I^{FS}=\frac{1}{2}\sum_{i\to N} \sum_{j\ne i} V\left(r_{ij}\right)-c\sum_i \left(\rho_i\right)^{1/2}\end{equation}
this is the Many-Body Finnis-Sinclair (FS) potentials developed with the intention to model the energetics of the transition metals. \textbf{\emph{They offer a better description of the surface relaxation in metals}}. They also avoid the problems,  such as the appearance of the Cauchy relation between the elastic constants $C_{12}=C_{44}$ which is not satisfied by cubic crystals, associated with using pair potentials to model metals.
\newline
where  \begin{equation}\rho_i=\sum_{j\ne i}\phi_{ij}\left(r_{ij}\right)\end{equation}
$V\left(r_{ij}\right)$ is a pairwise repulsive interaction between atoms $i$ and $j$, separated by a distance $r_{ij}$. Also, $\phi_{ij}\left(r_{ij}\right)$ are the two-body cohesive pair potentials, and $c$ is a positive constant. The second term in (3) is the cohesive many-body contribution to the energy.

\item  \begin{equation}H_I^{SC}=\varepsilon\left[\frac{1}{2}\sum_i \sum_{j\ne i} V\left(r_{ij}\right)-c\sum_i \left(\rho_i\right)^{1/2}\right]\end{equation}
this is the Many-Body Sutton-Chen (SC) Long-Range Potentials. They only describe the energies of 10 fcc elemental metals. They are similar to the FS type potentials and also similar to EAM potentials in form. SC potentials are very good for computer simulations of nanostructures involving large number of atoms, where $\varepsilon$ is a parameter with the dimensions of energy, 
\begin{equation}V\left(r_{ij}\right)=\left(\frac{a}{r_{ij}}\right)^n\end{equation}
\begin{equation}\rho_i=\sum_{j\ne i}\left(\frac{a}{r_{ij}}\right)^m\end{equation}
where $a$ is a parameter with the dimensions of length and usaully used to mean the equillibrium lattice constant. $m$ and $n$ are positive integers with $n>m$. The power law form was adopted so as to enable the model to be unique in combining the short-range interactions (due to the $N$-body second term of (5)) and good for describing the \textbf{\emph{surface relaxation}} phenomena using a van der Waals tail which gives a better description of the long-range interactions. 

\item  \begin{equation}U_{tot}=\sum_i \sum_{j > i} U_{ij}^{(2)}+\sum_i \sum_{j > i} \sum_{k > j} U_{ijk}^{(3)}\end{equation}
this is the Many-Body Murrel-Mottram (MM) Potentials. They are examples of cluster-type potentials and consist of sums of effective two- and three-body interactions. The pair interaction term, that is the first term, is modeled by a Rydberg function, which has been used for simple diatomic potentials. The first term takes the form below:
\begin{equation}\frac{U_{ij}^{(2)}}{D}=-\left(1+a_2 \rho_{ij}\right) exp\left(-a_2 \rho_{ij}\right)\end{equation}
where $\rho_{ij}=\frac{r_{ij}-r_e}{r_e}$. $D$ is the depth of the potential minimum, corresponding to the diatomic dissociation energy are $\rho_{ij}=0$, for $r_{ij}=r_e$, where $r_e$ is the diatomic equilibrium distance.
\begin{equation}\frac{U_{ijk}^{(3)}}{D}=P\left(Q_1,Q_2,Q_3\right)F\left(a_3, Q_1\right)\end{equation}
where 
\begin{eqnarray}P\left(Q_1,Q_2,Q_3\right)=c_0 + c_1Q_1 + c_2Q_1^2 + \nonumber\\
c_3\left(Q_2^2 + Q_3^2\right) + c_4Q_1^3 + c_5Q_1\left(Q_2^3+Q_3^2\right) +\nonumber\\ c_6\left(Q_3^3-3Q_3Q_2^2\right)\qquad\end{eqnarray}
this implies that there are seven unknowns in this model.
$F\left(a_3, Q_1\right)$ is the damping function and can take the following three forms:
\begin{eqnarray}F\left(a_3, Q_1\right)=exp\left(-a_3Q_1\right)\qquad exponential\nonumber\\
F\left(a_3, Q_1\right)=\frac{1}{2}\left[1-tanh\left(\frac{a_3Q_1}{2}\right)\right]\qquad tanh\nonumber\\
F\left(a_3, Q_1\right)=sech\left(a_3Q_1\right) \qquad sech\end{eqnarray}
The coordinates $Q_i$ are given as:
\begin{equation}\begin{bmatrix}Q_1\\Q_2\\Q_3\end{bmatrix}=\begin{vmatrix}\sqrt{1/3} & \sqrt{1/3} & \sqrt{1/3} \\ 0 & \sqrt{1/2} & -\sqrt{1/2} \\ \sqrt{2/3} & -\sqrt{1/6} & -\sqrt{1/6} \end{vmatrix} \begin{bmatrix}\rho_{ij}\\\rho_{jk}\\\rho_{ki}\end{bmatrix}\end{equation}

\item  \begin{eqnarray}U_{RTS}=\frac{1}{2}\sum_i \sum_{j \ne i} \hat{p_i}\hat{p_j}V_{AA}\left(r_{ij}\right)+
\left(1-\hat{p_i}\right)\left(1-\hat{p_j}\right)\times\nonumber\\
V_{BB}\left(r_{ij}\right)+ \left[\hat{p_i}\left(1-\hat{p_j}\right)+\hat{p_j}\left(1-\hat{p_i}\right)\right]V_{AB}\left(r_{ij}\right) -\nonumber\\
 d^{AA}\sum_i  \hat{p_i}\left[\sum_{i\ne j}\hat{p_j}\Phi_{AA}\left(r_{ij}\right)+
\left(1-\hat{p_j}\right)\Phi_{AB}\left(r_{ij}\right)\right]^{1/2}- \nonumber\\  
d^{BB}\sum_i\left(1-\hat{p_i}\right)\left[\sum_{i\ne j}\left(1-\hat{p_j}\right)\Phi_{BB}\left(r_{ij}\right)+
\hat{p_j}\Phi_{AB}\left(r_{ij}\right)\right]^{1/2} \end{eqnarray}
This is the Many-Body Rafii-Tabar-Sutton (RTS) Long-Range Alloy Potentials. They are, in particular, good for $A$-$A$, $A$-$B$ and $B$-$B$ fcc alloy metals. The RTS potentials are a form of generalization of the SC potentials and they model the energetics of the metallic fcc (face centered crystal) random binary alloys. They have the advantage of having little unknown parameters.
where 
\begin{equation}V_{\alpha\beta}\left(r\right)=\varepsilon^{\alpha\beta}\left[\frac{a^{\alpha\beta}}{r}\right]^{n_{\alpha\beta}}\end{equation}
\begin{equation}\Phi_{\alpha\beta}\left(r\right)=\left[\frac{a^{\alpha\beta}}{r}\right]^{m_{\alpha\beta}}\end{equation}
where $\alpha$, $\beta$ may be $A$ and $B$. The parameters $\varepsilon^{AA}$, $c^{AA}$, $a^{AA}$, $m^{AA}$ and $n^{AA}$ are for the pure element $A$, while the parameters $\varepsilon^{BB}$, $c^{BB}$, $a^{BB}$, $m^{BB}$ and $n^{BB}$ are for the pure element $B$.
we also have that
\begin{eqnarray}d^{AA}=\varepsilon^{AA}c^{AA}\nonumber\\ 
d^{BB}=\varepsilon^{BB}c^{BB}\end{eqnarray}
The mixed or pure alloy states or forms are contained in the following combining rules:
\begin{eqnarray}V_{AB}=\left(V^{AA}V^{BB}\right)^{1/2}\nonumber\\ 
\Phi_{AB}=\left(\Phi^{AA}\Phi^{BB}\right)^{1/2}\nonumber\\ 
m^{AB}=\frac{1}{2}\left(m^{AA}+m^{BB}\right)\nonumber\\ 
n^{AB}=\frac{1}{2}\left(n^{AA}+n^{BB}\right)\nonumber\\ 
a^{AB}=\left(a^{AA}a^{BB}\right)^{1/2}\nonumber\\ 
\varepsilon^{AB}=\left(\varepsilon^{AA}\varepsilon^{BB}\right)^{1/2}\end{eqnarray}
These parameters are very crucial to the MD simulations of elastic constants and heat of formation of a set of fcc metallic alloys as well as \textbf{\emph{for modelling the ultra thin Pd films on a Cu(100) surface}}. 
The operator $\hat{pi}$ is the site occupancy operator and is defined as:
\begin{eqnarray}\hat{pi}=1,\qquad if~the~site\nonumber\\ ~i~is~occupied~by~an~atom~A \nonumber\\ 
\hat{pi}=0, \qquad if~the~site~i~is\nonumber\\ ~occupied~by~an~atom~B\nonumber\\ 
\end{eqnarray}
\end{itemize}


\section{Methods}
\subsection{Making the potential general}
To make the potential general for studying this energy transport, delivered into a material by sputtering ion(s), we want to relate the following, listed, processes to one another. The reason being that they are related in sequential manner. It is our believe that these processes are needed in the energy transport caused by sputtering process.
\begin{enumerate}
\item the location of the landing ions will be called a sputtered location,
\item the origin of the nano-scale energy transport is from the, delivered, energy spread at the sputtered location,
\item all atoms captured by the energy spread will be regarded as the captured atoms,
\item it is important to note that there could be different atomic elements within the captured space,
\item the transportation of this energy causes ionizations of the captured atoms. Different atomic elements will receive and respond to this energy differently,
\item  We plan to apply the, rapidly changing, PAPCs (Photon Absorption Potential Coefficient) of each atom in that captured space to study the variations in the way each atom \textbf{\emph{receives}} energy, 
\item we have to note that even the same atoms, but located at different positions, in the sputtered location would become different due to PAPCs,
\item then we provide a way to evaluate the energy of the captured atoms in the, growing, sputtered location.
\end{enumerate}

Now, we concentrate on the development of two very crucial parameters: (1) the rate at which the number of ionised atoms increases and decreases due to the energy delivered by the sputtering particles landing on the surface and (2) the PAPC, which is essential in making the potential a general one and .

\subsection{Number of ionised atoms}
We could calculate the number of atoms that could be ionised by such energies as: 
\begin{equation}N_{ionised~atoms}=\frac{E_{ion}}{\sum_{i}^N I_i}\end{equation}
where $E_{ion}$ is the energy of the incoming ion and $I_i$ are the ionisation energies of the atoms in the lattice. For ionic lattices like that of NaCl, we could have two unique forms of $I_i$, that is: $I_\alpha$ and $I_\beta$. Therefore, equation (20) becomes
\begin{eqnarray}N_{ionised~atoms}=N_{\alpha~atoms}+N_{\beta~atoms}\nonumber\\ 
=\frac{E_{ion}}{\sum_{\alpha}^{N_{\alpha}} I_\alpha+ \sum_{\beta}^{N_{\beta}} I_\beta}\end{eqnarray}
Since we are looking at both laboratory and general cases of materials, we shall consider cases where the atomic lattice is very irregular (not necessarily amorphous) as well. In such cases (examples include a material made up of different homogenous layers of atoms), we have atomic arrangements where, say the $\alpha$ atoms are grouped together, more at a point, than at others. Ditto $\beta, \ldots$ atoms. For such complex molecules we have:
\begin{eqnarray}N_{ionised~atoms}=N_{\alpha~atoms}+N_{\beta~atoms} + \ldots \nonumber\\ 
=\frac{E_{ion}}{\sum_{\alpha}^{N_{\alpha}} I_\alpha+ \sum_{\beta}^{N_{\beta}} I_\beta+ \sum_{\ldots}^{N_{\ldots}} \ldots}\nonumber\\ 
\end{eqnarray}

Please note that we can not write (21 and 22) as :
\begin{equation}N_{ionised~atoms}=\frac{E_{ion}}{N_\alpha I_\alpha+ N_\beta I_\beta + \ldots}\end{equation} because
\begin{itemize}
\item even though the $\alpha$, $\beta$ and $\ldots$ atoms are fundamentally the same, they are electronically different at any point in time.
\item the direction or path along which the atoms receive energy must also be considered in determining how much of energy is left for subsequent ionisations.
\item the atoms might not be located close to one another at equal distances, 
\item it is also important that we discuss the fact that the population of ionised atoms is always changing in the lattice. This is because initially as the sputtering process starts, we shall have a larger number of first ionisation atoms but as the sputtering process increases, we have that number reducing while the number of nth ionisation atoms also increases. 
\item the point, just above, is what we have identified as a major reason why contributions to sputtering yield by temperature variations is crucial to a successful model for surface interactions. Research works on the contributions of temperature to sputtering yield is scarce and the very few model for such case is stated as follows:
\begin{equation}\lambda \sim \frac{1}{T^{1/2}}exp\left(-\frac{\bigtriangleup E}{2k_BT}\right)\end{equation} \textsuperscript{\cite{14}}
where $\lambda$ is the ripple wavelength, $T$ is the temperature, $\bigtriangleup E$ is the activation energy of the system and $k_B$ is the Boltzman constant.
\end{itemize}

Let us consider, for now, two groups of ionisations: first and second ionisations. The first ionisation is handled as :
\begin{equation}N_{first~ionised~atoms}=N_1=1-tanh(\xi) \end{equation}
while the second as :
\begin{equation}N_{second~ionised~atoms}=N_2=tanh(\xi)\end{equation}
as $\xi$ increases, the $N_1$ reduces and $N_2$ increases. However, the equations above are similar to:
\begin{equation}N_{second~ionised~atoms}=N - N_1=N_2\end{equation}
where N is the total number of atoms in the lattice, assuming that we can only have two (first and second) ionisations.

In general, for a complex system, we have:
\begin{eqnarray}N_{first~ionised~atoms}=N_{ij}\nonumber\\ 
=N_1\left(\delta_{ij}+\frac{\left(-1\right)^{ij}}{2}tanh(\xi)\right)\end{eqnarray}
\begin{eqnarray}N_{second~ionised~atoms}=N_{jk}\nonumber\\ 
=N_2\left(\delta_{jk}+\frac{\left(-1\right)^{jk}}{2}tanh(\xi)\right)\end{eqnarray}
\begin{eqnarray}N_{third~ionised~atoms}=N_{kl}\nonumber\\ 
=N_3\left(\delta_{kl}+\frac{\left(-1\right)^{kl}}{2}tanh(\xi)\right)\end{eqnarray}
\newline
$\ldots$
\newline
\begin{eqnarray}N_{nth~ionised~atoms}=N_{mn}\nonumber\\ 
=N_{nth}\left(\delta_{mn}+\frac{\left(-1\right)^{mn}}{2}tanh(\xi)\right)\end{eqnarray}

So we have:
\newline
$N_{ij}=N_1+N_2\Rightarrow N_2=N_{ij}-N_1$
\newline
$N_{jk}=N_2+N_3\Rightarrow N_3=N_{jk}-N_2$
\newline
$N_{kl}=N_3+N_4\Rightarrow N_4=N_{kl}-N_3$
\newline
$\ldots$
\newline
$\therefore$
\newline
\begin{equation}N_3=N_{jk}-\left(N_{ij}-N_1\right)=N_{jk}-N_{ij}+N_1\end{equation}
\begin{equation}N_4=N_{kl}-N_{jk}+N_{ij}-N_1\end{equation}
\begin{equation}N_5=N_{lm}-N_{kl}+N_{jk}-N_{ij}+N_1\end{equation}
$\ldots$
\newline
Please note that $N_4$,$N_5$,$N_6$ etc population will be $\approx 0$ at the initial stages of the sputtering process. As time factor $\xi$ grows, $N_4$,$N_5$,$N_6$ will start emerging while $N_1$,$N_2$ start disappearing.

Now, we can re-write equation (22) as:
\begin{eqnarray}N_{ionised~atoms}=\nonumber\\ 
\frac{E_{ion}}{N_{pq} - \ldots N_{lm}-N_{kl}+N_{jk}-N_{ij}+N_1}\end{eqnarray}
\newline
\begin{itemize}
\item Please note that $\xi$ is the condition of time frame during the whole process of sputtering,
\item Also, we shall define $\delta_{ij}$, $\delta_{jk}$, $\ldots$ as the conditions for the transfer of ionisation from one stage to another,
\item the $\frac{1}{2}$ in equations (28) to (31) is for the purpose of linking one stage of ionisation to the other,
\item the $N_1$, $N_2$, $\ldots$ in equations (28) to (31) are the initial numbers of each stage of ionisation.
\end{itemize}

\subsection{Introducing the PAPC}
PAPC is our way of finding out the way similar (such as two Na atoms but at different ionisations) or different (such as Na and Cl atoms at the same or different ionisations) atoms would react to the energy delivered by the sputtering ion(s). Therefore, we arrive at the following: 
\begin{eqnarray}A_c=\frac{1}{2 z}\left\{\ln\left(\frac{1+z^-}{1-z^-}\right)+\ln\left(\frac{1+z^+}{1-z^+}\right)\right\}\nonumber\\
=\frac{1}{2 z}\left\{\ln\left(\frac{1+z^- +z^+ +z^-\times z^+}{1-z^-  -z^+ +z^-\times z^+}\right)\right\}\end{eqnarray}
Equation(36) can also be written as:
\begin{equation}A_c=\frac{1}{z}\left( \tanh^{-1}(z^-) + \tanh^{-1}(z^+)\right)\end{equation}
where $z^-$ is is given as:
\begin{equation}z^-=\frac{z_v^-}{\left|z-z_v^-\right|}\end{equation}
and $z^+$ is is given as:
\begin{equation}z^+=\frac{z_v^+}{\left|z-z_v^+\right|}\end{equation}
where $z_v^+$ and $z_v^-$ are the extra subtracted and added electrons of the neutral electrons of the atom in question. Take for instance, the $A_c$ of an atom like $Na$ is will be given as:
\begin{itemize}
\item for first ionisation, we have:
$z_v^+ = \frac{12}{\left|11-12\right|}=12$ and $z_v^- = \frac{10}{\left|11-10\right|}=10$. $A_c=0.0167148$
\item for second ionisation, we have:
$z_v^+ = \frac{13}{\left|11-13\right|}=6.5$ and $z_v^- = \frac{9}{\left|11-9\right|}=4.5$. $A_c=0.0346427$
\item for third ionisation, we have:
$z_v^+ = \frac{14}{\left|11-14\right|}=4.66667$ and $z_v^- = \frac{8}{\left|11-8\right|}=2.66667$. $A_c=0.0556262$ 
\end{itemize}

\subsection{Derivations for the general potential general}
To further study this perturbation, we shall apply the principle that the energy thrown off by the, gradually, approaching foreign particle captures and causes an ionization of a group of atoms on the surface. This, group, ionization leads to a build up of forceful repulsive energy amongst the captured atoms. This repulsion eventually translates to surface sputtering and penetration of the incoming particle. Therefore, this incoming particle's energy does work on the surface of the material it approaches. To this end, we introduce the proposed work done on the surface by the incoming beams of ions. Before we state the proposed work done, we state the repulsion potential suitable for our work done, comparable to those well known potentials such as Lennard-Jones' and Morse's \textsuperscript{\cite{11}},\textsuperscript{\cite{12}}, as follows:
\begin{equation}\Upsilon=\gamma\left( e^{\left(\frac{1}{\pi\sigma}\right)} - \frac{\hat{\kappa}\hat{\mu}}{\gamma^2}\right)\Gamma\left(\zeta\right)\end{equation}
where 
\begin{itemize}
\item $\gamma$ is the energy of the incoming ion, 
\item $\sigma\to\sigma_{i+1}=\sigma_i+\bigtriangleup\sigma$, [where $\bigtriangleup\sigma=exp\left(-\frac{\gamma}{\theta\sigma}\right)$] is the distance covered by the photon cone of the ion. If the sputtering angle, $\theta$ which can take values from 0 and 90 degrees, is $0$, that is the approach on the surface is perpendicular to the surface, then we have $\bigtriangleup\sigma=0$.
\item \footnotesize\begin{eqnarray*}\hat{\kappa}=\frac{k~region}{\sum~atomic~diameters~of~atoms~in~k~region}\qquad\qquad\nonumber\\
=\frac{k~region}{\sum D_{pq} - \ldots \sum D_{lm}-\sum D_{kl}+\sum D_{jk}-\sum D_{ij}+\sum D_1}
\end{eqnarray*}
\end{itemize}
\begin{itemize}
\item is the \textbf{\emph{lattice site occupation density}} of the topograph of the surface in which the ion is to land. Please note that the default $k$ region for a flat surface is a semi-circular shape. Where $\sum D_{pq}$ is the sum of all atoms in the $N_{pq}$ population. Ditto others.
\item \begin{equation}\hat{\mu}=A_c \times atomic~number\end{equation} is the property of the molecules on whom the ion is to land. The properties are the PAPC and the atomic numbers of the atoms of that molecule, 
where PAPC is given as earlier stated.
\begin{eqnarray}A_r=A_c\frac{e}{R}\nonumber\end{eqnarray}
where $e$ is the electronic charge and $R$ is the increasing distance of the transporting energy from the landing site of the foreign particle.
\item \begin{equation}\zeta=\beta\hat{\mu}\sigma\end{equation} is the quantity that determines the activation of the sputtering as the ion is approaching the surface. $\beta$ is a constant that stabilizes the rate of increment or reduction of $\sigma$,
\item \begin{equation}\Gamma\left(\zeta\right)=1-tanh\left(\zeta\right)\end{equation} helps to activate the sputtering process. This is because when the ion is still far off, above the surface, there will not be sputtering but as it approaches the surface, the sputtering process starts to build, gradually. Please see figure 1 for illustrations.
\end{itemize}

Therefore, we may wish to write:
\begin{equation}\Upsilon=\gamma\left( e^{\left(\frac{1}{\pi\sigma}\right)} - \frac{\hat{\mu}\int d^3 \sigma}{\gamma^2 \sum_i D_i}\right)\Gamma\left(\zeta\right)\end{equation}
where $D_i$ is the diameters of the atoms in the k region.

Finally, we now have the proposed work done on the surface, by one ion's photon shower, to be given as:
\begin{equation}\Omega=\sigma\frac{d\Upsilon}{d\sigma}\end{equation}
\begin{eqnarray}\Omega=\gamma exp\left(\frac{1}{\pi\sigma^2}\right) -\frac{\sigma\hat{\mu}\partial_{\sigma}\int d^3\sigma}{\gamma^2\sum_i D_i} -
\nonumber\\
\left(\gamma exp\left(\frac{1}{\pi\sigma^2}\right)-\frac{\sigma\hat{\mu}\partial_{\sigma}\int d^3\sigma}
{\gamma^2\sum_i D_i}\right)tanh\left(\beta\hat{\mu}\sigma\right)-
\nonumber\\
\beta\hat{\mu}\sigma\left(\gamma exp\left(\frac{1}{\pi\sigma^2}\right)+\frac{\hat{\mu}\partial_{\sigma}\int d^3\sigma}{\gamma^2\sum_i D_i}\right)sech^2\left(\beta\hat{\mu}\sigma\right)
\end{eqnarray}
We present the plots of our derived repulsion potential, for some selected (flat) surfaces at $\gamma=$ 10eV, $k$ region is $\frac{2}{3}\pi\sigma^3$ and $\beta=$ 0.5, as a way to compare with the parameters of the repuslion potentials such as Lennard-Jones' and Morse's.


\section{Results and conclusion}
Now, we present the results of the simulation of equations (1) and (9) in the following diagrams. Figures 1 and 2 illustrate how energy interactions with the surface are in phases, real sputtering is actived, gradually, as it gets closer to the surface. The circular shapes are the, default, gradual energy spreads. Figures 3 to 8 are plots of the repulsive potentials against the repulsion distances $\sigma$. They are comparism between the literature (Lennard-Jones' and Morse's potentials) and our proposed potential. In all the figures, $\gamma$=10eV and $k$=$\frac{2}{3}\pi\sigma^3$. We do not consider the mixing ratios of the alloys and so our figures 3 to 9 do not have any specific molecular formula. Tables 1, 2 and 3 show more comparison between this paper and the literature on the results of simulations carried out on other sampled materials. Figures 9 and 10 are the plots of phase shift in the perturbed molecular dynamics of the materials considered. They show that all materials have frequently changing (perturbed) molecular dynamics as it experiences surface bombardments. In figure 10, the two plots suggest, using our study of the perturbation of the molecular dynamics in materials as, a (probable) reason why gold does not rust as easily as iron. It our conclusion that it is best if these small perturbations are not ignore so as to fine tune and accurately channel the energy delivery mechanisms. This will be undoubtedly required in medical applications where cross-organ energy transport is undesirable.  Furthermore, going by our results obtained, we have clearly shown that the MD perturbations in a material, via its surface, is peculiar to the surface. 

\begin{multicols}{2}
\begin{center}
\includegraphics[width=0.45\textwidth, height=5cm]{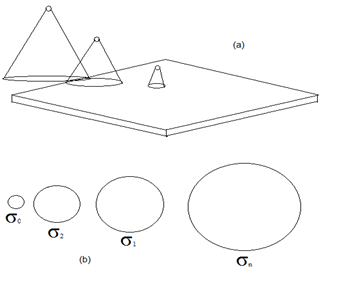}
\end{center}
\captionof{figure}{}

\begin{center}
\includegraphics[width=0.45\textwidth, height=5cm]{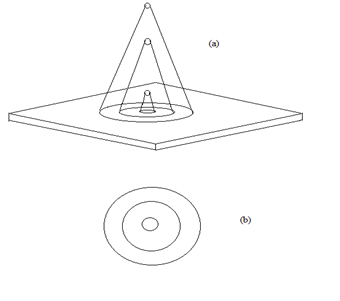}
\end{center}
\captionof{figure}{}

\begin{center}
\includegraphics[width=0.45\textwidth, height=5cm]{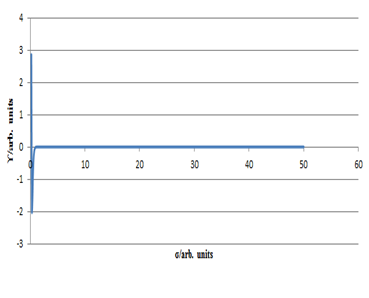}
\end{center}
\captionof{figure}{}

\begin{center}
\includegraphics[width=0.45\textwidth, height=5cm]{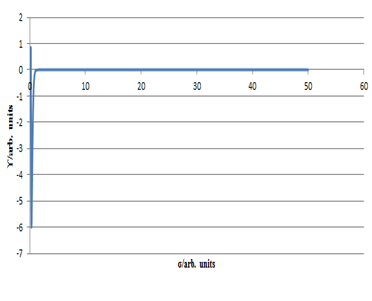}
\end{center}
\captionof{figure}{}

\begin{center}
\includegraphics[width=0.45\textwidth, height=5cm]{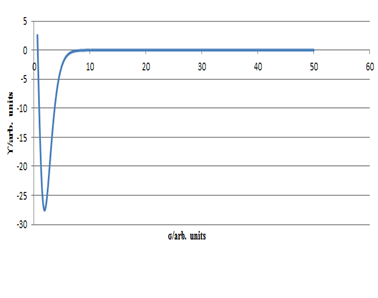}
\end{center}
\captionof{figure}{}

\begin{center}
\includegraphics[width=0.45\textwidth, height=5cm]{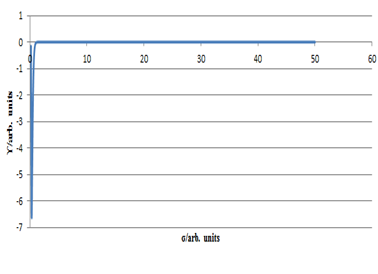}
\end{center}
\captionof{figure}{}

\begin{center}
\includegraphics[width=0.45\textwidth, height=5cm]{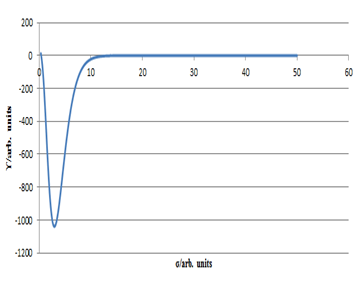}
\end{center}
\captionof{figure}{}

\begin{center}
\includegraphics[width=0.45\textwidth, height=5cm]{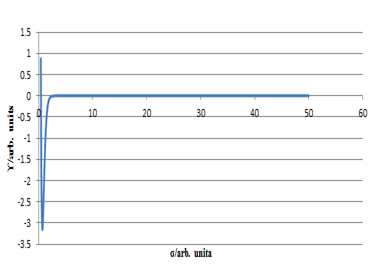}
\end{center}
\captionof{figure}{}

\captionof{table}{}
\begin{tabular}{l*{3}{c}r}
\toprule
Potential		& 	Value			&	$\sigma/nm$	\\
\midrule
Morse's		&	-0.942869		&	3		\\
Lennard-Jones' 	&	-0.0110927		&	3		\\
our potential		& 	-17.2711		&	98(2)		\\
\bottomrule	
\end{tabular}

\captionof{table}{}
\begin{tabular}{l*{3}{c}r}
\toprule
\multicolumn{3}{c}{ Morse's Potential.} \\
\midrule
Material		& 	 Potential/(eV)	&	$\sigma/nm$	\\
Tungsten		&	-0.989553		&	2		\\
Zinc			&	-0.154966		&	2		\\
Copper		& 	-0.327924		&	2		\\
\toprule
\multicolumn{3}{c}{ Our Potential.} \\
\midrule
Material		& 	 Potential/(eV)	&	$\sigma/nm$		\\
Tungsten		&	-16746.0		&	98(2)	\\
Zinc			&	-3916.38		&	95(5)	\\
Copper		& 	-4812.23		&	95(5)	\\
\bottomrule	
\end{tabular}

\captionof{table}{}
\begin{tabular}{l*{3}{c}r}
\toprule
\multicolumn{3}{c}{Lennard-Jones' Potential} \\
\midrule
Material		& 	 Potential/(eV)		&	$\sigma/nm$		\\
$SF_6$		&	-207.007		&	5		\\	
$CS_2$		&	-402.091		&	4		\\
$NH_3$		& 	-244.099		&	3		\\
$H_2O$		&	-502.195		&	2		\\
$CH_3CN$		&	-486.32		&	4		\\
$Si\left(CH_3\right)_4$	& 	-313.775	&	6		\\
\toprule
\multicolumn{3}{c}{ Our Potential} \\
\midrule
Material		& 	 Potential/(eV)		&	$\sigma/nm$	\\
$SF_6$,$\beta=0.0003$		&	-38.2094		&	98(2)	\\	
$CS_2$,$\beta=0.0003$		&	-38.3425		&	98(2)	\\
$NH_3$,$\beta=0.003$		& 	-23.2576		&	96(4)	\\
$H_2O$,$\beta=0.003$		&	-1.88068		&	98(2)	\\
$CH_3CN$,$\beta=0.003$	&	-3.30931		&	97(3)	\\
$Si\left(CH_3\right)_4$,$\beta=0.0003$		& 	-218.492		&	96(4)	\\	
\bottomrule	
\end{tabular}

\end{multicols}

\begin{center}
\includegraphics[width=0.95\textwidth, height=5cm]{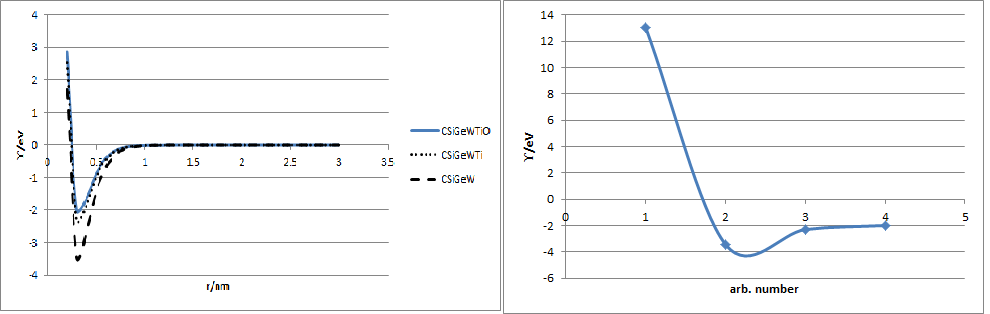}
\end{center}
\captionof{figure}{}

\begin{center}
\includegraphics[width=0.95\textwidth, height=5cm]{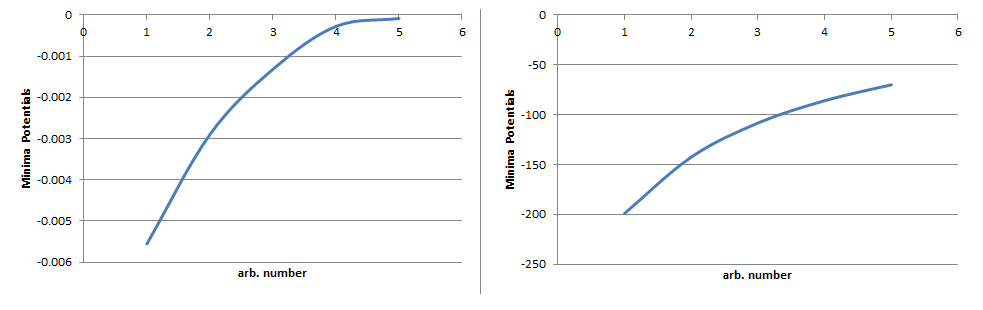}
\end{center}
\captionof{figure}{}

\vspace*{0.5in}
List of captions
\newline\newline
Figure 1: The same ion approaching the surface at slant angles.
\newline\newline
Figure 2: The same ion at normal to the surface.
\newline\newline
Figure 3: Repulsion potential of equation (6) for CSiGeWTiO. The equilibrium position is at $\sigma=$ 4. This is a case of quintenary alloy surface.
\newline\newline
Figure 4: Repulsion potential of equation (6) for CSiW. The equilibrium positions are at $\sigma=$ 4, 4, 4.1 and 3.8 respectively. This is a case of tertiary alloy surface.
\newline\newline
Figure 5: Repulsion potential of equation (6) for CSiO. The equilibrium positions are at $\sigma=$ 4, 4, 4.1 and 3.8 respectively.
\newline\newline
Figure 6: Repulsion potential of equation (6) for WTiO. The equilibrium positions are at $\sigma=$ 4, 4, 4.1 and 3.8 respectively. These are another cases of tertiary alloy surface.
\newline\newline
Figure 7: Repulsion potential of equation (6) for CSiGe. The equilibrium positions are at $\sigma=$ 4, 4, 4.1 and 3.8 respectively. This is another case of tertiary alloy surface.
\newline\newline
Figure 8: Repulsion potential of equation (6) for TiO. The equilibrium position is at $\sigma=$ 3.8. This is a case of binary alloy surface.
\newline\newline
Figure 9: $\textbf{left~ image:}$ Potential plot for CSiGeWTiO, CSiGeWTi and CSiGeW. $\textbf{right ~image:}$ Phase shift in the molecular dynamics of the changing potential in the binding process from CSiGe to CSiGeW to CSiGeWTi and to CSiGeWTiO at the same distance 0.3nm starting with CSiGe.
\newline\newline
Figure 10: $\textbf{left ~image:}$  Plot of the minimum (equilibrium) potential for transformation from $Au$ to $Au + O$ to $Au + 2O$ to $\ldots$ to $Au + 5O$. $\textbf{right ~image:}$  Plot of the minimum (equilibrium) potential for transformation from $Fe$ to $Fe + O$ to $Fe + 2O$ to $\ldots$ to $Fe + 5O$.
\newline\newline
Table 1: Minimum values for each of the figures. Comparing the results of Morse's, Lennard-Jones' and our potential for Argon.
\newline\newline
Table 2: Comparing \textbf{left:} Morse's with \textbf{right:} our potential, $\beta=0.00003$.
\newline\newline
Table 3: Comparing \textbf{left:} Lennard-Jones' with \textbf{right:} our potential





\end{document}